\documentclass{article}
\usepackage{ismir}
\usepackage[T1]{fontenc} 
\usepackage[utf8]{inputenc} 
\usepackage{amsmath,cite,url}
\usepackage{graphicx}
\usepackage{float}

\usepackage{color}
\usepackage{algpseudocode}

\usepackage[firstpage]{draftwatermark}
\definecolor{lightgray}{rgb}{0.9,0.9,0.9}
\definecolor{darkgray}{rgb}{0.4,0.4,0.4}
\SetWatermarkFontSize{12pt}
\SetWatermarkScale{1.1}
\SetWatermarkAngle{90}
\SetWatermarkHorCenter{202mm}
\SetWatermarkVerCenter{170mm}
\SetWatermarkColor{darkgray}
\SetWatermarkText{Late-Breaking / Demo Session Extended Abstract ISMIR 2022 Conference}

\def\lbd{}	

\title{Audio Latent Space Cartography}

\twoauthors
  {Nicolas Jonason} {KTH Royal Institute of Technology \\ {\tt njona@kth.se}}
  {Bob L. T. Sturm} {KTH Royal Institute of Technology \\ {\tt bobs@kth.se}}

\def\authorname{N Jonason, BLT Sturm}

\usepackage[bookmarks=false,pdfauthor={\authorname},pdfsubject={\papersubject},hidelinks]{hyperref}

\begin{document}

\maketitle
\begin{abstract}
We explore the generation of visualizations of audio latent spaces using an audio-to-image generation pipeline. We believe this can help with the interpretability of audio latent spaces. We demonstrate a variety of results on the NSynth dataset.
\end{abstract}

\section{Introduction}
\begin{figure}[ht]
\centering
 \includegraphics[width=\columnwidth]{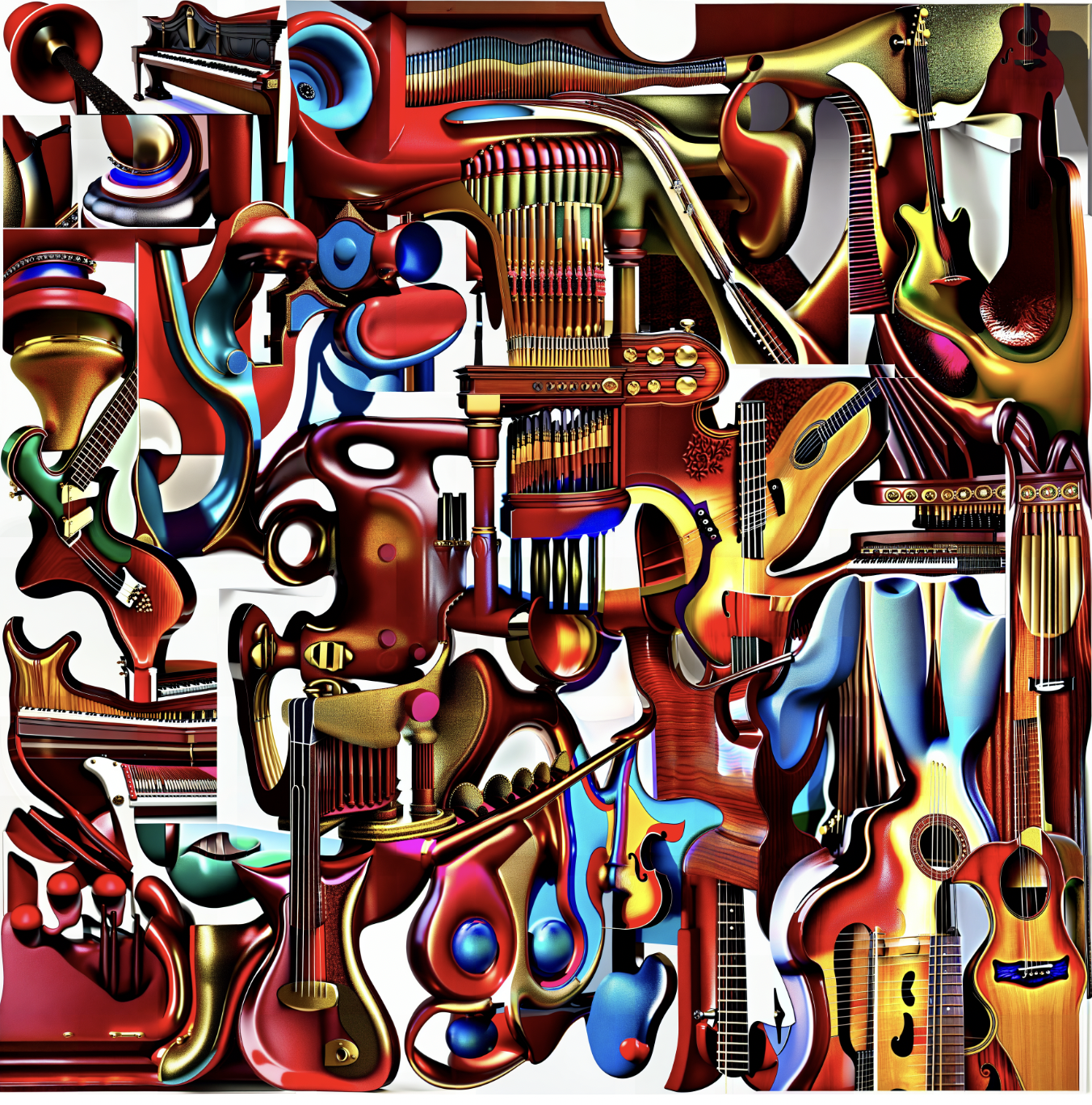}
 \caption{Map of a latent space containing 60 sounds from the NSynth dataset}
 \label{fig:map}
\end{figure}

Many techniques exist for visualizing high dimensional latent spaces \cite{liu_latent_2019}.
We present a new technique for providing visual support in the exploration of audio latent spaces.
At a high level, this technique works by drawing audio samples from the latent space and using a audio-to-image generation pipeline to generate images which are then combined to form a map of the latent space. 
A web demo is available.\footnote{\url{https://erl-j.github.io/audio-latent-space-cartography-demo}}
\section{Approach}

\begin{figure}[!ht]
    \centering
    \includegraphics[width=\columnwidth]{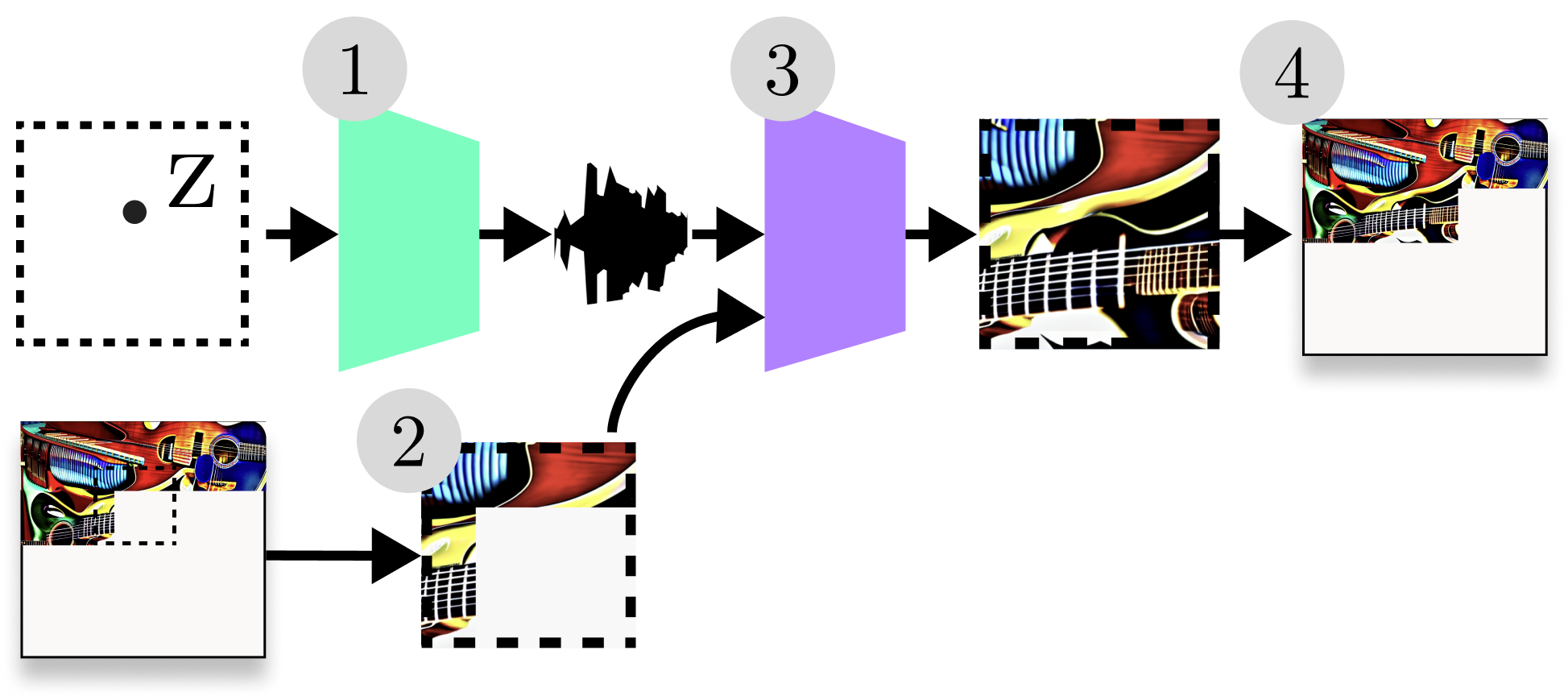}
    \caption{Diagram showing an iteration of the audio latent space map creation process. For each iteration, we draw latent coordinates in order to: 1) Generate audio; 2) Extract a patch of the current map to be inpainted. Then, we use an audio-to-image generation pipeline (TimbreCLIP + Stable Diffusion) to inpaint the patch (3). Finally, the map is updated with the inpainted patch (4)}
    \label{fig:explanation}
\end{figure}

Figure \ref{fig:explanation} shows the audio latent space map creation process.
The process involves the TimbreCLIP + Stable Diffusion audio-to-image pipeline described in \cite{jonason_timbreclip_2022}. 
At a high level, this pipeline interpolates the embeddings of multiple prepared text prompts based on how closely the input waveform relates to various keywords. 
The interpolated prompt embedding is then used to guide Stable Diffusion.\cite{rombach_high-resolution_2022}
The keywords and template used to generate prompts are a crucial hyper-parameter which allow us to control how the audio latent space is depicted. Please refer to \cite{jonason_timbreclip_2022} for details about the audio-to-image generation pipeline.

\begin{figure}
\centering
 \includegraphics[width=1.0\columnwidth]{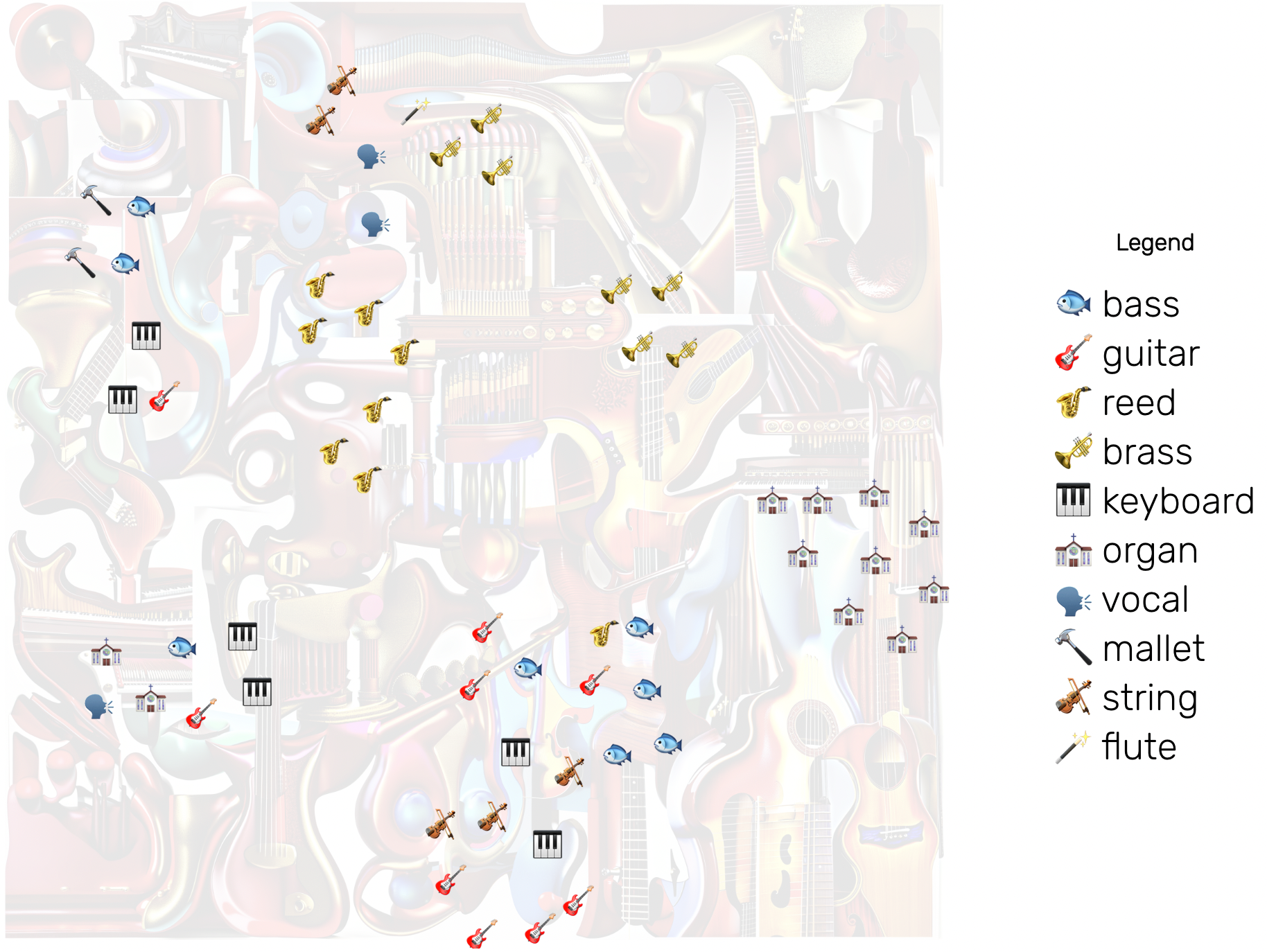}
 \caption{Latent map of 60 sounds from NSynth showing the coordinates of the 60 sounds in the latent space as well as their respective instrument families.}
 \label{fig:mapwlegend}
\end{figure}

\begin{figure}
    \centering
    \includegraphics[width=1.0\columnwidth]{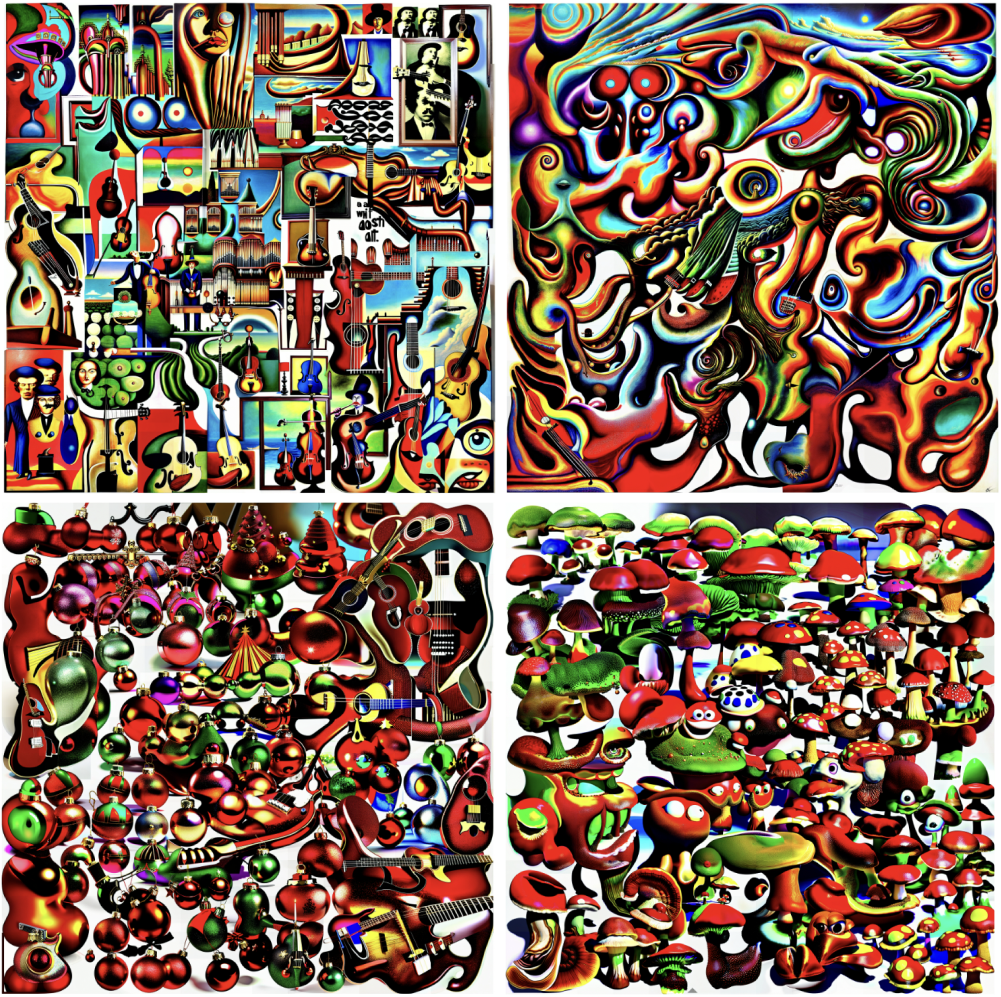}
    \caption{4 maps of the same audio latent space in different styles. The map in the top left uses the style of the artist René Magritte, top right uses the style of the artist Zdzisław Beksiński. The map in the bottom left is in the style of Christmas ornaments, bottom right map depicts the audio latent space as mushrooms. Keywords and templates for these examples and more are available in the web demo.}
    \label{fig:styles}
\end{figure}

\section{Demonstration}
Figure \ref{fig:map} shows a map created with our method. The latent space itself was constructed by applying the UMAP \cite{mcinnes_umap_2020} algorithm to TimbreCLIP embeddings \cite{jonason_timbreclip_2022} of sounds from the NSynth dataset \cite{engel_neural_2017}. 
We use 60 sounds from 10 different instrument families. All sounds were produced by playing the MIDI pitch C4. Figure \ref{fig:mapwlegend} shows the same latent space map with the position and labels of the 60 samples superimposed onto the latent map. 

Due to time constraints, we opted for the following simplification of the approach described in the previous section: instead of transforming every coordinate and then using that audio to drive the image generation, we instead generate a TimbreCLIP embedding using the inverse UMAP transformation and feed the TimbreCLIP embedding directly into the image synthesizer.

The prompt template used in the audio-to-image generation pipeline for Figures \ref{fig:map} and \ref{fig:mapwlegend} is \texttt{"A 3D rendered close-up of a <KEYWORD>, pinterest trending aesthetic"}.
The keywords are a list of names of 21 musical instruments:
\texttt{"bass guitar",
"acoustic guitar",
"piano keyboard",
"flute",
"pipe organ",
"violin",
"cello",
"double bass",
"violin",
"viola",
"saxophone",
"trumpet",
"trombone",
"tuba",
"clarinet",
"marimba",
"kalimba",
"xylophone",
"bell",
"electric guitar",
"human voice"}.
Maps using different prompt templates and keywords are showcased in Figure \ref{fig:styles} as well as in the web demo. 

\section{Acknowledgments}

This paper is an outcome of MUSAiC, a project that has
received funding from the European Research Council
under the European Union’s Horizon 2020 research and
innovation program (Grant agreement No. 864189).

\clearpage
\bibliography{references}

%
%
%
%
%

\end{document}